\documentclass[prc,aps,twocolumn,groupedaddress,showpacs,floatfix]{revtex4}
\usepackage{graphicx}
\usepackage[usenames]{color}

\usepackage{amsbsy}

\newcommand{\beq}{\begin{equation}}
\newcommand{\eeq}{\end{equation}}
\newcommand{\bea}{\begin{eqnarray}}
\newcommand{\eea}{\end{eqnarray}}

\newcommand{\benn}{\begin{displaymath}}
\newcommand{\eenn}{\end{displaymath}}

\begin{document}

\title{Spin-polarized low-density neutron matter}

\author{Alexandros Gezerlis}
\affiliation{Department of Physics, University of Washington, Seattle, Washington, 98195, USA}

\begin{abstract}
Low-density neutron matter is relevant to the study of neutron-rich nuclei and neutron star crusts.  
Unpolarized neutron matter has been studied extensively over a number of decades, while experimental
guidance has recently started to emerge from the field of ultracold atomic gases. 
In this work, we study population-imbalanced neutron matter (possibly relevant to magnetars and to
density functionals of nuclei)  
applying a Quantum Monte Carlo method that has proven successful 
in the field of cold atoms. We report on the first \textit{ab initio} 
simulations of superfluid low-density polarized neutron matter.
For systems with small imbalances, we find a linear dependence of the energy 
on the polarization, 
the proportionality coefficient being dependent on
the density. We also present results for the momentum and pair distributions of the two fermionic components.
\end{abstract}

\date{\today}


\maketitle

\section{Introduction}

The inner crust of a neutron star is widely considered to be composed of 
a lattice of neutron-rich nuclei along with a gas of neutrons and electrons.
The gas of neutrons is expected to be superfluid at weak to intermediate
coupling. Thus, low-density neutron matter is intrinsically connected to
the strongly coupled fermion many-body problem, necessitating accurate 
calculations (or simulations). 
Low-density neutron matter is of relevance to the static and dynamic 
properties of the neutron star crust,
which can lead to observable behavior. \cite{Page:2009,Aguilera:2009,Fattoyev:2010}
Outside the observational realm, neutron matter computations also hold 
significance in the context of traditional nuclear physics:
equation of state results at densities close to the nuclear saturation density
have been used for some time to constrain  Skyrme and other density functional approaches 
to heavy nuclei, while the density-dependence of the $^1\mbox{S}_0$ gap in low-density neutron matter 
has also been used to constrain Skyrme-Hartree-Fock-Bogoliubov treatments 
in their description of neutron-rich nuclei \cite{Chamel:2008}.
The potential significance of such calculations has led to a series of publications on the equation
of state of low-density neutron matter over the last few decades.
\cite{Friedman:1981,Akmal:1998,Carlson:Morales:2003,Schwenk:2005,Margueron:2007b,Baldo:2008,Epelbaum:2008b,Epelbaum:2008a,Abe:2009,Gandolfi:2009,Rios:2009,Hebeler:2010} 

Parallel developments in a separate field of physics have recently provided new insight as well as the promise
of direct experimental constraints: experiments with ultracold atomic gases of fermions are now being carried out in a number
of labs around the world. In some cases, atomic gases in traps contain sufficiently many particles that the local-density
approximation is valid. As a result, such experiments can measure the energy \cite{Luo:2009,Navon:2010} and pairing gap 
\cite{Carlson:2008,Schirotzek:2008} of homogeneous strongly interacting matter.
For cold fermionic atoms, the two-particle interaction can be directly tuned using a magnetic field through so-called Feshbach resonances 
to produce a specific scattering length $a$, while the effective range $r_e$ between the atoms is 
considerably smaller than the average interparticle distance, and thus essentially zero. 
These conditions are analogous to low-density neutron matter, where the 
particle-particle interaction has a scattering length which is
considerable, $\approx -18.5$ fm, and is therefore larger than the average
interneutron spacing. On the other hand, 
cold atoms and low-density neutron matter are clearly distinct systems: 
first, for neutron matter the effective range is much smaller than the scattering length,
$r_e \approx 2.7$ fm, so $ |r_e / a| \approx 0.15$, but only at very low densities is the
effective range much smaller than the interparticle spacing. Second, the neutron-neutron (NN) interaction
is not strictly limited to $s$-waves, implying a complicated spin dependence. Third, three-neutron
interactions (NNN) are in principle also present. The last two points can be remedied if one studies 
relatively low-densities, i.e. at most an order of magnitude smaller than the nuclear
saturation density. The first point might be addressed in the framework of cold atomic experiments in the future:
it may be possible to use narrow and wide resonances in cold atoms to study
this experimentally.\cite{Marcelis:2008}

The above discussion is limited to unpolarized neutron matter, i.e. to the case of
two species of particles, conventionally called spin-up ($\uparrow$) and spin-down ($\downarrow$), 
with equal populations. However, the general case of imbalanced systems has also been studied 
extensively: limiting ourselves for the moment to neutron matter, calculations of the magnetic
susceptibility have been appearing consistently since the discovery of pulsars. \cite{Haensel:1975}
Given that the magnitude of the pairing gap is approximately $1$ MeV, a magnetic field in a neutron star crust 
would have to be larger than $10^{16}-10^{17}$ G to polarize neutron matter. The question of spin-polarized
neutron matter is thus in principle relevant to the objects known as magnetars, which have surface magnetic
fields of $10^{14}-10^{15}$ G. Thus, the promise of observational insight into these objects has led a number
of theoretical groups to study spin-polarized neutron matter and the associated question of a possible
ferromagnetic instability at large density. 
\cite{Fantoni:2001,Vidana:2002,Margueron:2002,Rios:2005,Bordbar:2007,Sammarruca:2007,Perez:2009}

In this work, we take a step back and address neutron matter with a finite spin-polarization 
(population imbalance) \textit{at low density}. This system is in principle relevant to neutron star observations:
in a realistic neutron-star crust polarization may appear at lower densities than for
infinite matter. For example, the spin-orbit splitting around a large nucleus
might help favor polarization at a lower magnetic field than
would be required for the bulk.
Furthermore, our results could also be used as an input to or benchmark for phenomenological theories of terrestrial nuclei.
On a different note, as already mentioned, the region of low-density is close to the physics of cold atoms, since there the NN interaction
is simpler and the NNN interaction is minimal. Thus, we can use approaches already verified in the laboratory
with ultracold atomic gases. Importantly, we use a Quantum Monte Carlo (QMC) method which has been applied, in
previous works, to polarized cold atomic systems, \cite{Carlson:2005,Pilati:2008} as well as 
to unpolarized cold atoms and neutron matter \cite{Carlson:2003,Gezerlis:2008,Gezerlis:2010}.
Our earlier works were consistent both with experimental measurements and with the analytically
known behavior of the energy and the gap at vanishingly small coupling.\cite{Lee:1957,Gorkov:1961}
We extend this QMC method appropriately, allowing us to study spin-polarized superfluid neutron matter
and therefore provide the first benchmark calculation of this system using an \textit{ab initio} microscopic 
simulation approach.

In such systems, it is conventional to use the following measure of the population imbalance:
\begin{equation}
P = \frac{N_{\uparrow}-N_{\downarrow}}{N_{\uparrow}+N_{\downarrow}}
\end{equation}
where $N_{\uparrow}$ and $N_{\downarrow}$ are the numbers of spin-up and spin-down particles, respectively, and
$P$ is called the polarization. The regime of large polarization is related to a question
that has a long history in the framework of the BCS theory. In the BCS approach, superfluidity arises from the pairing 
of particles of different spin occupying states of opposite momenta near the Fermi surface. 
In the case of spin imbalance, the Fermi surfaces of the two components
no longer coincide, making pairs with zero total momentum difficult to form. At some finite polarization, the gap between
the two Fermi surfaces becomes so large that the system undergoes a quantum phase transition to a normal state 
(this is known as the Chandrasekhar-Clogston limit).

Our aim in this work is to provide quantitatively reliable results for superfluid low-density neutron matter. 
We therefore limit our simulations to small polarizations, $P < 0.1$, exploring the regime where 
pairing is likely to be energetically stable. We assume there are no exotic superfluid phases and no phase
separation. We calculate ground-state energies at 
different total number densities ($\rho = (N_{\uparrow} + N_{\downarrow})/L^3$),
more specifically at $\rho_1 = 6.65 \times 10^{-4}$, $\rho_2 = 2.16 \times 10^{-3}$, and $\rho_3 = 5.32 \times 10^{-3}$ fm$^{-3}$. 
To put these densities into perspective we can compare them to nuclear matter
saturation density: they are $0.41, 1.35$, and $3.32$ percent, respectively, of $\rho_0 = 0.16$ fm$^{-3}$.
At each total density, we study the cases of $35+33$, $37+33$, and $39+33$ particles (see below).
We also compute the momentum distributions and pair-distribution functions for the two different components.

\section{Quantum Monte Carlo}
\label{sec:qmc}
\subsection{Hamiltonian}
\label{sec:hamilt}

As pointed out in the Introduction, we do not need to include NNN interactions, 
since we are interested in a density regime where
these are quite small.
Thus, we use the following non-relativistic Hamiltonian:
\begin{equation}
{\cal{H}} = \sum\limits_{k = 1}^{N}  ( - \frac{\hbar^2}{2m}\nabla_k^{2} )  
+ \sum\limits_{i,j} v_4(r_{ij}) ~,
\label{ham}
\end{equation}
where $N = N_{\uparrow}+N_{\downarrow}$ is the total number of particles. 
The full neutron-neutron interaction is complicated, 
having one-pion exchange at large distances, an intermediate 
range spin-dependent attraction by two-pion exchange, and a short-range repulsion. 
As already discussed, however, in dilute neutron matter the 
dominant contributions come from the opposite-spin pairs, and specifically 
from the scattering length and the effective range, along with a short-range repulsive 
core which is important so as to avoid a collapse to a higher-density state

In this work we are including an excess of neutrons of one species. Thus, it is also significant to take
into account the same-spin interactions in Eq. (\ref{ham}), which we do by using
the interaction introduced in Ref. \cite{Gezerlis:2010}.
This interaction includes a propagator (see next subsection) in which 
all opposite-spin pairs interact through the $^{1}\mbox{S}_0$ channel of the 
Argonne v18 (AV18) \cite{Wiringa:1995} potential, which fits $s$-wave nucleon-nucleon scattering
very well at both low- and high-energies. 
Thus, in what follows, for the purposes of the evolution the spins are considered
to be ``frozen'', with the majority species being called $\uparrow$ and the minority
species being called $\downarrow$.
We explicitly include the $p$-wave interactions in the same-spin pairs, 
and perturbatively correct the $S=1, M_S=0$
pairs to the correct $p$-wave interaction. We use the Argonne v4$'$ (AV4$'$) potential to determine 
the $p$-wave interactions.\cite{Wiringa:2002}.

Since we're studying neutrons, the AV4$'$ interaction can be written as follows:
\begin{equation}
v_4(r) = v_c(r) + v_{\sigma}(r){\mbox{\boldmath$\sigma$}}_1\cdot{\mbox{\boldmath$\sigma$}}_2,
\label{vfour}
\end{equation}
which in the case of the $S=0$ (singlet) pairs gives:
\begin{equation}
v_S(r) = v_c(r) - 3v_{\sigma}(r)~.
\label{ves}
\end{equation}
In turn, the contribution from the $S=1$ (triplet) pairs has the form:
\begin{equation}
v_P(r) = v_c(r) + v_{\sigma}(r)~.
\end{equation}
The same-spin potential contribution is small, but as the population imbalance 
increases the relative weights also change accordingly (see section \ref{sec:eos}). 
While still keeping the potential of Eq. (\ref{ves}) in the propagator of our QMC method for the opposite-spin pairs,
we have introduced a perturbative correction by writing Eq. (\ref{vfour}) in terms
of the Majorana exchange operator, which exchanges
the positions leaving the spins unaffected:
\begin{equation}
v_4(r) = v_c(r) + v_{\sigma}(r)(-2P^M - 1)
\label{vmajorana}
\end{equation}

\subsection{Variational and Green's Function Monte Carlo}
\label{sec:gfmc}
The first step in our microscopic simulation is a Variational Monte Carlo 
(VMC) calculation. Variational Monte Carlo is a relatively simple combination of classical
Monte Carlo and the variational (Rayleigh-Ritz) method; it was
first used by McMillan in the 1960s.
It is based on a variational trial
wave function $\Psi_{V}$ that is a reasonably good approximation of the true ground-state wave
function. It contains variational parameters that in principle should allow one to approach
the true wave function (see the next subsection for more details).
A VMC calculation uses Monte Carlo integration
to minimize the expectation value of the Hamiltonian:
\begin{equation}
\langle H \rangle_{VMC} = \frac{\int d{\bf R} \Psi_{V}({\bf R}) H \Psi_{V}({\bf R})}{\int d{\bf R} |\Psi_{V}({\bf R})|^2} \geq E_0~,
\label{eq:qmcvmc}
\end{equation}
thus optimizing the variational wave function $\Psi_V$. The fact that it is relatively easy to
perform a VMC simulation allows us to examine various possibilities in placing the excess particles
in different momentum states.

It is customary to use the output configurations of a Variational Monte Carlo calculation
as input to a more extensive calculation using the method known as Green's Function
Monte Carlo (GFMC). This method works by projecting out the exact, lowest-energy
eigenstate $\Psi_{0}$ from a trial (variational) wave function $\Psi_{V}$ by treating the
Schr\"{o}dinger equation as a diffusion equation in imaginary time $\tau$ and stochastically evolving the
variational wave function for a ``sufficiently'' long time. 

The evolution operator $e^{-iHt}$ becomes $e^{-H\tau}$ in imaginary time, 
commonly written as $e^{-(H-E_T) \tau}$, where the $E_T$ is called the trial energy. 
Applying this operator to the variational wave function and expanding in terms of the
complete set of eigenstates gives:
\begin{eqnarray}
\Psi(\tau) & = &
 e^{-(H-E_T) \tau} \Psi_V
  =  \sum_i \alpha_i e^{-(E_i - E_T) \tau} \Psi_i \nonumber \\
 &=& \alpha_0 e^{-(E_0-E_T) \tau} \Psi_0, \hspace{1em} \lim \tau \rightarrow \infty~.
\label{evolution}
\end{eqnarray}
The GFMC technique is implemented by discretizing $\tau$ and expressing the imaginary-time propagator as
\begin{equation}
e^{-(H-E_T)\tau} = \prod\limits_{n} e^{-(H-E_T)\triangle\tau}~,
\end{equation}
where $\tau = n \triangle\tau$. If we now define the short-time Green's function by:
\begin{equation}
G({\bf R},{\bf R}^{\prime}) = \langle{\bf R}|e^{-(H-E_T)\triangle\tau}|{\bf R}^{\prime}\rangle~,
\end{equation}
where ${\bf R}$ is the configuration vector ${\bf R} = ({\bf r}_{1},{\bf r}_{2} \ldots {\bf r}_{N})$ of $3N$ dimensions, 
then we can use it to calculate the evolved $\Psi(\tau)$ starting from a set of VMC configurations.
The short-time Green's function 
can be conveniently approximated using the Trotter-Suzuki formula:
\begin{eqnarray}
G({\bf R},{\bf R}^{\prime}) &\approx& e^{-V({\bf R})\frac{\triangle\tau}{2}} \langle{\bf R}|e^{-T\triangle\tau}|{\bf R}^{\prime}\rangle e^{-V({\bf R}^{\prime})\frac{\triangle\tau}{2}} e^{E_T\triangle\tau} \nonumber \\
&=& e^{-(V({\bf R})+ V({\bf R}^{\prime}) - 2E_T)\frac{\triangle\tau}{2}} \langle{\bf R}|e^{-T\triangle\tau}|{\bf R}^{\prime}\rangle~,
\label{short}
\end{eqnarray}
which is accurate to order $(\Delta \tau)^2$.

To avoid the fermion-sign problem, we impose what is known as the
``fixed-node approximation''. A fixed-node simulation leads to
a wave function $\Psi_0$ that is the lowest-energy state with the same
nodes as the trial wave function $\Psi_V$. The resulting
energy $E_0$ is an upper bound to the true ground-state energy.
Thus, if one chooses the variational wave function so that it includes 
a number of parameters, \cite{Carlson:2003} these
parameters can be optimized to give the best approximation 
to the ground-state wave function (see next subsection).

\subsection{Trial wave function}
\label{sec:wave}

In these VMC and GFMC calculations there is a need to express the wave function of the system in terms of 
specific coordinate-space states. To this effect, we use a finite number $N$ of particles with Born-von Karman (periodic) 
boundary conditions in a cubic box of volume $L^3$, and $N$ is chosen to be large enough so that the system 
can be assumed to be in the thermodynamic limit. For neutron matter, this was shown to be approximately 66 particles in Ref. \cite{Gezerlis:2008}. 
Using a Cartesian coordinate system, the quantized plane waves $e^{ i {\bf{k}}_{\bf n} \cdot {\bf r}}$ will have momentum vectors of 
the following discrete form:
\begin{equation}
{\bf k}_{\bf n} = \frac{2 \pi}{L} ( n_{x}, n_{y}, n_{z})~,
\label{bornagaineq}
\end{equation}
where the $n_x, n_y, n_z$ are integers. The shell number $I$ is defined such that $I = n_x^2 +n_y^2 + n_z^2$. 
Thus, there is only 1 possible combination of the $n_x, n_y, n_z$ that gives $I=0$, 6 combinations that produce $I=1$, 
12 combinations that lead to $I=2$ and so on. Neutrons are spin one-half fermions, therefore, for equal populations
for the two components the system has 
a closed-shell structure when $N = 2, 14, 38, 54, 66, \ldots$.

The simplest possible approximation (which, strictly speaking, is applicable only to the case of closed shells) that can be used 
for the input variational wave function is to describe the particles as being in a free Fermi gas.
This approach assumes no correlations in the wave function and is equivalent to having a product of two Slater determinants, 
one for spin-up and one for spin-down:
\begin{equation}
\Phi_S({\bf{R}}) = D_{\uparrow} D_{\downarrow}.
\label{slater}
\end{equation}
The single-particle states in the Slater determinants are $\phi_n(r_k) = e^{ i {\bf{k}}_{\bf n} \cdot {\bf r}_k}/L^{3/2}$. 

Another choice for $\Phi({\bf{R}})$, one which can also describe pairing, is the well-known BCS wave function $\Phi_{BCS}({\bf{R}})$ 
in its form for fixed particle number (which reduces to the Slater case under specific conditions). 
This choice is agreeable for both physical reasons (it reflects the 
fact that fermions with an attractive interaction can form Cooper pairs in the ground state) and mathematical reasons ({\itshape unlike} 
the Slater wave function, it has nodal surfaces which can be varied so as to minimize the fixed-node GFMC energy). Furthermore, a 
computationally appealing aspect of this wave function is the fact that it can be 
written down as a determinant. \cite{Bouchaud:1988}

In this formalism, a general wave function with $n$ pairs, $u$ spin-up and $d$ spin-down 
unpaired particles can be written as: 
\begin{widetext}
\begin{equation}
\Phi_{\rm BCS}({\bf{R}}) =  {{\cal A}} \left \{ 
\left[ \phi(r_{11'})...\phi(r_{nn'})\right] \left [
\psi_{1\uparrow}({\bf r}_{n+1})...\psi_{u\uparrow}({\bf r}_{n+u})\right ]
\left[\psi_{1\downarrow}({\bf r}_{(n+1)'})...\psi_{d\downarrow}({\bf r}_{(n+d)'})\right ]
\right \} ~. 
\end{equation}
The unpaired particles are placed in $\psi_{i\uparrow}$ and $\psi_{j\downarrow}$ single-particle states. 
We can write this wave function as the determinant of an
$M\times M$ matrix where $M= n+u+d $.
In this work, we are interested in the case of a ``gapless superfluid''
which for polarized neutrons translates to $33$ opposite-spin pairs
along with an excess of unpaired particles of one species.
When we have $2$ extra spin-up particles, the corresponding matrix
is written as follows:
\begin{equation}
\left (
\begin{array}{cccccc}
\phi(r_{1,1'}) & \phi(r_{1,2'}) & ... &
\phi(r_{1,33'}) &
\psi_{1\uparrow}({\bf r}_1) & \psi_{2\uparrow}({\bf r}_1) \\
\phi(r_{2,1'}) & \phi(r_{2,2'}) & ... &
\phi(r_{2,33'}) &
\psi_{1\uparrow}({\bf r}_2) & \psi_{2\uparrow}({\bf r}_2) \\
\vdots & \vdots & \vdots\vdots\vdots & \vdots & \vdots & \vdots\\
\phi(r_{35,1'}) & \phi(r_{35,2'}) & ... &
\phi(r_{35,33'}) &
\psi_{1\uparrow}({\bf r}_{35}) & \psi_{2\uparrow}({\bf r}_{35}) \\
\end{array}
\right )
\end{equation}
\end{widetext}

The pairing function
$\phi (r)$ is a sum over the momenta compatible with the periodic boundary conditions. In the BCS theory the pairing function is:
\begin{equation}
\phi(r) =\sum\limits_{\bf n} \frac{v_{{\bf k}_{\bf n}}}{u_{{\bf k}_{\bf n}}} e^{i{\bf k}_{\bf n}\cdot{\bf r}} =\sum_{\bf n} \alpha_n e^{i{\bf k}_{\bf n}\cdot{\bf r}} ~,
\end{equation}
and here it is parametrized with a short- and long-range part as in Ref. \cite{Carlson:2003}:
\begin{equation}
\phi({\bf r}) = \tilde{\beta} (r) + \sum_{{\bf n},~I \leq I_C} \alpha_I e^{ i {\bf{k}}_{\bf n} \cdot {\bf r}}~,
\label{phieq}
\end{equation}

We choose the single-particle states, $\psi_{i\uparrow}$, to be plane waves so as to ensure momentum conservation.
We pick their momentum by checking values near the minimum (at each density) of the quasiparticle dispersion. 
The latter is calculated using the odd-even energy staggering:
\begin{equation}
\Delta = E(N+1) - \frac{1}{2} \left [ E(N)+E(N+2) \right ]~,
\label{eq:staggerer}
\end{equation}
where $N$ is an even number of particles. At each density, the minimum of the dispersion lies at a different momentum.\cite{Gezerlis:2010} 
As already mentioned, we used VMC to place the particles at different momentum states.
For very small polarizations, the minimum system energy is expected to be identical to the minimum of the dispersion, which follows
from adding only one extra particle. This is indeed the result we find, the only exceptions appearing at
density $\rho_3$ and particle numbers of $39+33$ and higher (see below).

In practice, we also include Jastrow (correlation) terms in the variational wave function:
\begin{equation}
\Psi_V = \prod\limits_{i \neq j} f_P(r_{ij}) \prod\limits_{i' \neq j'} f_P(r_{i'j'}) \prod\limits_{i,j'} f(r_{ij'})  \Phi_{\rm BCS}({\bf{R}})
\end{equation}
where the unprimed (primed) indices refer to spin-up (spin-down) particles.
The Jastrow parts are taken from a lowest-order-constrained-variational method \cite{Pandharipande:1973} calculation described 
by a Schr\"{o}dinger-like equation:
\begin{equation}
- \frac{\hbar^2}{m}\nabla^{2} f(r)  + v(r) f(r) = \lambda f(r)~\nonumber
\end{equation}
for the opposite-spin $f(r)$ and 
\begin{equation}
- \frac{\hbar^2}{m}\nabla^{2} f_P(r)  + v(r) f_P(r) + \frac{2\hbar^2}{m r^2} f_P(r) = \lambda f_P(r)~\nonumber
\end{equation}
for the same-spin $f_P(r)$. Since the $f(r)$ and $f_P(r)$ we use are nodeless, they do not affect the final result 
apart from reducing the statistical error. 
Since we are using the fixed-node approximation,
we know that the result we obtain for one set of pairing function parameters 
in Eq. (\ref{phieq}) will be an upper bound to the true ground-state energy
of the system. The 
parameters are optimized in the full GFMC calculation as in previous works \cite{Carlson:2003, Gezerlis:2008}, providing the
best possible nodal surface, in the sense of lowest fixed-node energy, 
for that form of trial function.
As mentioned in section \ref{sec:gfmc}, this upper-bound property 
allows us to get as close as possible to the true ground-state energy of the spin-polarized superfluid system.

\section{Results}
\label{sec:res}
\subsection{Equation of state}
\label{sec:eos}

We first address the energy of spin-polarized low-density neutron matter versus polarization.
We have studied three total densities $\rho_1 = 6.65 \times 10^{-4}$, $\rho_2 = 2.16 \times 10^{-3}$, 
and $\rho_3 = 5.32 \times 10^{-3}$ fm$^{-3}$. 
A smaller density would correspond to neutron matter that
is closer to the neutron star surface, which in turn implies a smaller magnetic field, and is thus less
likely to be polarized. Also, lower density is more difficult
to propagate in imaginary time to a satisfactory accuracy level, given that lower density
leads to larger inverse energy and therefore longer propagation. 
Reversely, we do not study even larger densities because then we would have
to include 3-body interactions in our approach. For pure neutron matter this would imply, first, going away 
from firm experimentally constrained interactions (three-neutron interactions are commonly fit to $N=Z$ nuclei) 
and, second, the necessity of using spin-isospin
dependent wave functions, therefore disallowing the use of approximately 70 particles and thus the simulation of the
thermodynamic limit in the framework of a variational \textit{ab initio} approach. 
Furthermore, larger densities would imply that the afore-mentioned perturbative
correction in the propagator (see section \ref{sec:hamilt}) would break down. For these calculations to be quantitatively reliable, 
the number of opposite-spin pairs should not stray too much from the ``canonical'' $33+33$ case.
Thus, for each of the three densities, we address the cases of $35+33$, $37+33$, and $39+33$ particles. 
For the case of the largest density 
we have also examined $41+33$ and $43+33$ particles (i.e. up to nearly double the polarization),
finding the same overall trend.

\begin{table}[t]
\caption{Results for the ground-state energy divided with the total number of particles
for two species of neutrons interacting via the AV4$'$ potential at $\rho_3 = 5.32 \times 10^{-3}$ fm$^{-3}$.}
\begin{center}

\begin{tabular}{c c c c}
\hline
 $N_{\uparrow}+N_{\downarrow}$ & $E_{full}$ [MeV] & $E_{\uparrow\uparrow}$ [keV] & $E_{\downarrow\downarrow}$ [keV]\\
\hline
33+33 & 2.133(1) & 11.4(1) & 11.3(1) \\
35+33 & 2.178(2) & 12.2(1) & 10.7(1) \\
37+33 & 2.230(2) & 12.7(1) & 10.0(1) \\
39+33 & 2.286(2) & 13.8(2) & 9.6(1) \\
\hline
\end{tabular}

\end{center}
\label{table1}
\end{table}

In most works on polarized neutron matter, the dependence of energy on polarization is taken
to be quadratic.\cite{Fantoni:2001,Vidana:2002,Margueron:2002,Rios:2005,Bordbar:2007,Sammarruca:2007,Perez:2009} 
However, these works address dense neutron matter, in which the $^1\mbox{S}_0$ pairing has
already reached gap closure, in the presence of a strong magnetic field. 
In other words, most works in the literature study normal matter, in which 
an isolated spin-flip only impacts particles near their respective Fermi surfaces. 
In this work we study low-density polarized neutron matter, implying that
superfluidity plays an important role: flipping a spin 
is equivalent to breaking a pair. Thus, 
the pairing gap plays a decisive role in producing the polarized state.
Our results for the energy versus polarization exhibit a linear trend:
\begin{equation}
\frac{E}{N}(\rho,P) = \frac{E}{N}(\rho,0) + \alpha(\rho) P
\end{equation}
similarly to the case of ultracold atomic gases at unitarity \cite{Carlson:2005},
where Quantum Monte Carlo calculations found a linear dependence of the energy on
polarization at small population imbalances.

In this connection, it is relevant to examine the interaction between (polarized) quasiparticles.
Summing single-particle excitation energies (taken from Ref. \cite{Gezerlis:2010}) for particles that are
placed at the appropriate minimum momentum in Eq. (\ref{bornagaineq}) we find
that the full Quantum Monte Carlo results are thus approximately reproduced. This
implies that the quasiparticles are weakly interacting.

\begin{figure}[t]
\vspace{0.5cm}
\begin{center}
\includegraphics[width=0.46\textwidth]{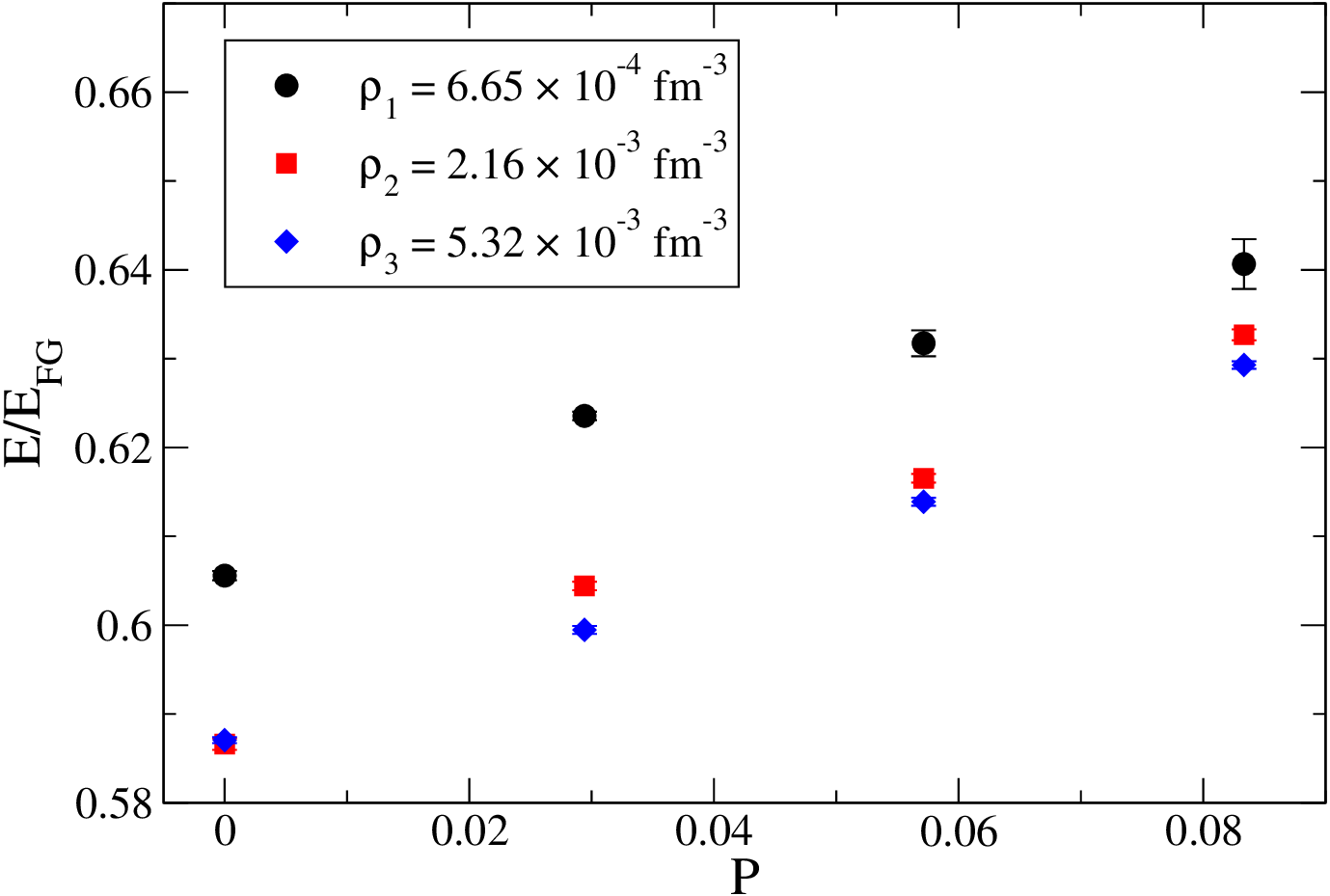}
\caption{(color online) 
Ground-state energy per particle (in units of the free Fermi gas energy) for spin-polarized neutron matter. 
Shown are QMC results at three different total densities.
The overall trend is linear, the slope depending on the density.}
\label{fig:ener}
\end{center}
\end{figure}

The $\alpha(\rho)$ coefficients we have extracted from these results 
for the three densities are $0.37(5), 1.01(5)$, and $1.84(9)$ MeV, respectively. 
In Fig. \ref{fig:ener} we show the energy per particle at different densities versus polarization. 
To facilitate comparison between results at different densities we have
divided the energy per particle with the energy of a free Fermi gas at the same total density:
\begin{equation}
E_{FG} = \frac{3}{10} \frac{\hbar^2}{m} (3\pi^2 \rho)^{2/3}~.
\end{equation}
We notice that, just like in the case of unpolarized neutron matter \cite{Gezerlis:2010},
when the density increases the energy in units of $E_{FG}$ drops, but the rate of the drop is also decreasing.
The slight deviation from linear behavior at $\rho_1$ stems from the  afore-mentioned necessity 
to propagate  up to longer imaginary times (at least by a factor of 3) in comparison to the other
cases. This is also the reason why the results for the bigger systems at that density have
larger error bars.

Further results for the energy of polarized neutron matter are given in Table \ref{table1}, which refers to the largest density we have studied, $\rho_3 = 5.32 \times 10^{-3}$ fm$^{-3}$. 
At this density we find the maximum value of the perturbative correction, which is 9 percent of the total energy.
As should be expected, for the
$33+33$ system the energies of the $\uparrow \uparrow$ and the $\downarrow \downarrow$
interactions are identical (within statistical error). As we increase the polarization,
there is a clear trend toward the increase of the $\uparrow \uparrow$ relative importance,
which for the $43+33$ system becomes nearly double the corresponding
$\downarrow \downarrow$ contribution.
In all cases, the same-spin contribution to the energy is small, not growing to more than 1 percent of the total
energy.

\subsection{Distribution functions}
\label{sec:distrib}

We have also used GFMC to calculate distribution functions at $\rho_3 = 5.32 \times 10^{-3}$ fm$^{-3}$ for
37 $\uparrow$ and 33 $\downarrow$ particles: 
in contradistinction to the case of the energy,
the results for these functions are not upper bounds to the true ground-state results, but they are expected to be accurate 
(the error being of second order in $|\Psi_0 - \Psi_V|$).

\begin{figure}[t]
\vspace{0.5cm}
\begin{center}
\includegraphics*[width=0.46\textwidth]{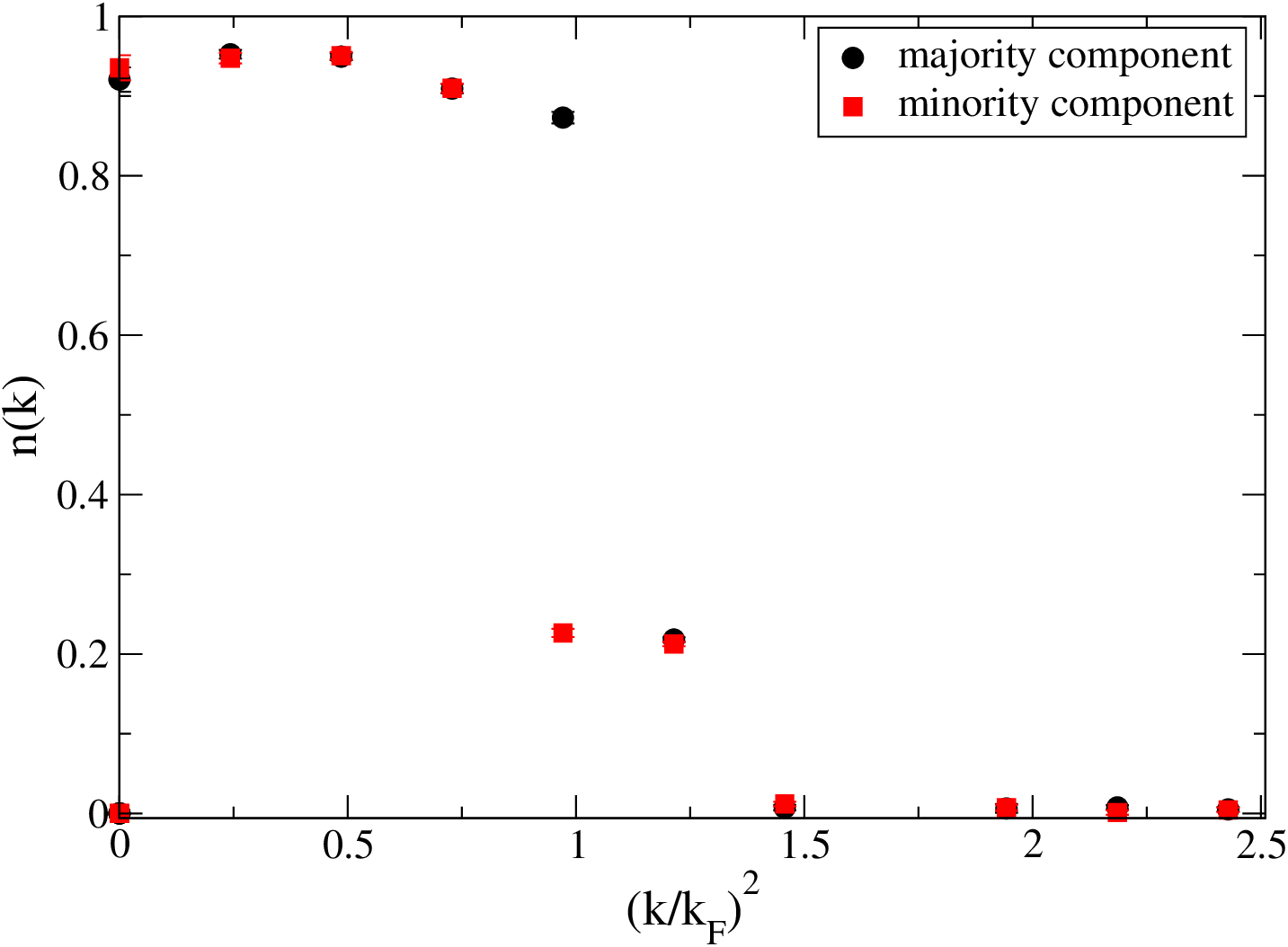}
\caption{(color online) 
Momentum distribution versus $(k/k_F)^2$ 
for the two different spins, for the case of 
37 $\uparrow$ and 33 $\downarrow$ particles 
(total density $\rho_3 = 5.32 \times 10^{-3}$ fm$^{-3}$), shown as
circles and squares respectively. $k_F$ is taken here to refer to the
total density, $k_F = \left (3 \pi^2 \rho_3 \right)^{1/3}$.
The behavior exhibited is commonly referred to as ``Fermi surface mismatch''.}
\label{fig:modi}
\end{center}
\end{figure}

Starting with the momentum distribution, we first discuss our expectations using mean-field BCS theory as a guide.
BCS is not quantitatively accurate in this regime, \cite{Gezerlis:2010} but can provide
qualitative understanding. In BCS, the momentum distribution is given by the following expression:
\begin{equation}
 n({\bf k}) = \frac{1}{2} \left [ 1 - \frac{\xi({\bf k})}{E({\bf k})} \right ]~,
\end{equation}
where $\xi({\bf k}) = \epsilon({\bf k})-\mu$, the chemical potential is $\mu$ and $\epsilon({\bf k}) = \frac{\hbar^2k^2}{2m}$ 
is the single-particle energy of a particle with momentum ${\bf k}$. The elementary quasi-particle excitations of the system have energy:
\begin{equation}
E({\bf k}) = \sqrt{\xi({\bf k})^2+\Delta({\bf k})^2}
\label{quasienereq}
\end{equation}
Overall, this is close to a step function for small gaps, but it changes considerably in the strong 
coupling regime. In general, the spread of the momentum distribution around $\mu$ is approximately 
$2 \Delta$. At this density, the system exhibits a gap of $\Delta = 1.05(11)$ which as 
a fraction of the Fermi energy is $\Delta/E_F = 0.17(2)$ implying that there is no clearly defined Fermi surface. \cite{Gezerlis:2010}
Even so, in the case of spin-polarized Fermi gases it is customary to use the language of weak coupling
and speak of a ``Fermi surface mismatch''. This follows from the fact that the Fermi energy is proportional
to $\rho_{{\uparrow}({\downarrow})}^{2/3}$ and in this case the densities for the two spin populations are different. 

In Fig. \ref{fig:modi} we show the momentum distribution computed using GFMC.
This is calculated as the Fourier transform of the one-body density matrix, through:
\begin{equation}
n_{\uparrow(\downarrow)}(k) \equiv \frac{N_{\uparrow(\downarrow)}}{L^3} 
\left\{ \int d \delta r e^{i {\bf k} \cdot ({\bf r}_{n}' - {\bf r}_{n})} 
\frac{\Psi_{V}({\bf r}_1, \ldots, {\bf r}_n')}{\Psi_{V}({\bf r}_1, \ldots, {\bf r}_n)} \right\}~,
\end{equation}
where the curly brackets denote a stochastic integration over the angles. The integral 
over $\delta r = |{\bf r}_{n}' - {\bf r}_{n}|$ is performed on a line analytically to avoid statistical 
errors due to the oscillatory radial dependence. In both cases, we see a considerable spread
around the chemical potential value, but we also notice a clear distinction in how the two species
behave around that point, the majority species showing a ``lag'' in its decline. 

\begin{figure}[t]
\vspace{0.5cm}
\begin{center}
\includegraphics*[width=0.46\textwidth]{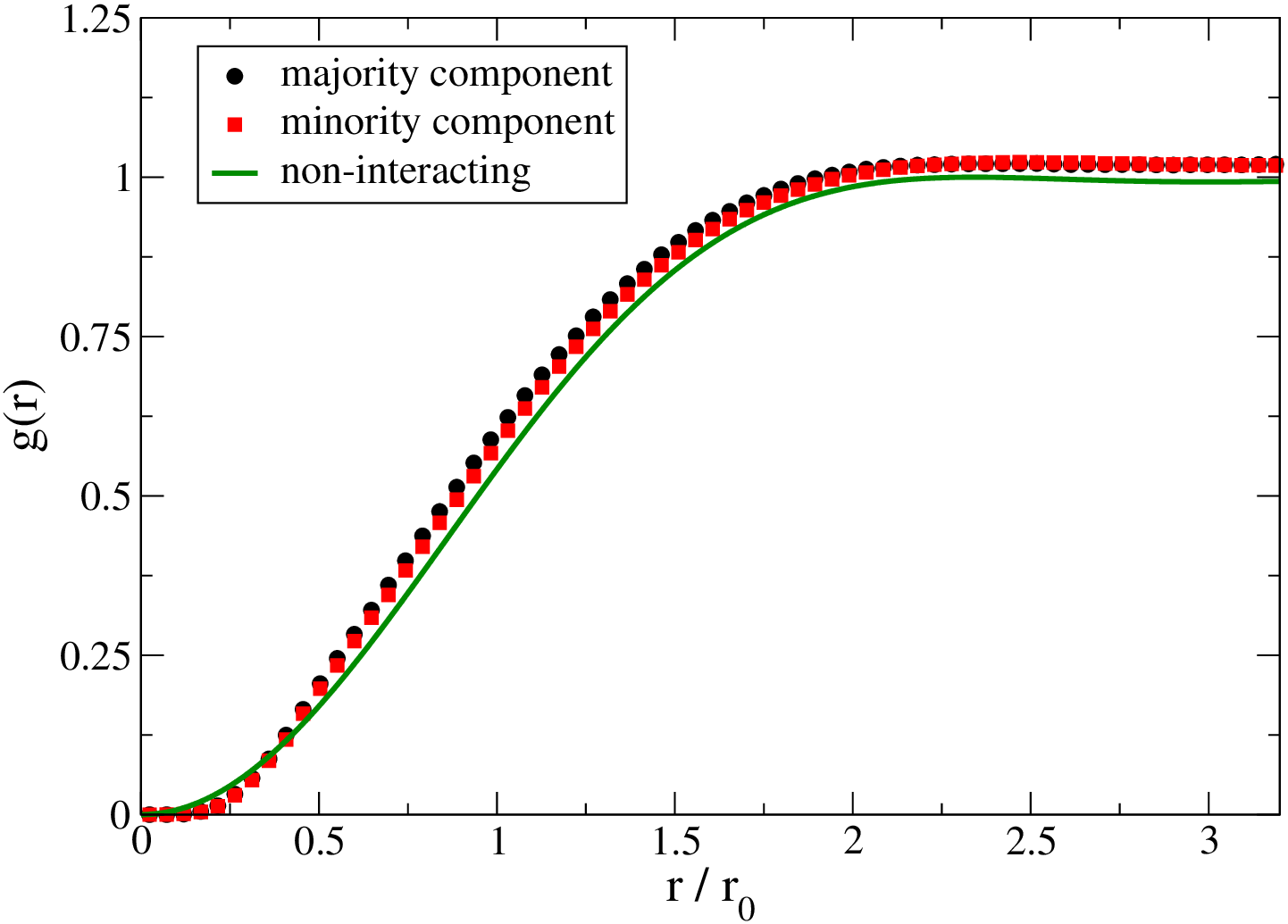}
\caption{(color online) 
Same-spin pair-distribution function as a function of the distance divided with a measure
of the average interparticle spacing,  
for the case of 37 $\uparrow$ and 33 $\downarrow$ particles (total density $\rho_3 = 5.32 \times 10^{-3}$ fm$^{-3}$), shown as
circles and squares respectively. Also given is the same-spin pair-distribution function for a 
non-interacting Fermi gas (line).}
\label{fig:padi}
\end{center}
\end{figure}

We have also computed the pair-distribution functions and have plotted 
them in Fig. \ref{fig:padi}. These are calculated from an expectation value of the form:
\begin{equation}
g_P(r) = A \sum\limits_{i<j} \langle \Psi_0 | \delta(r_{ij}-r) O^P_{ij} | \Psi_V \rangle~,
\label{eq:CApadi}
\end{equation}
where we are interested in the case in which the operator is simply unity, and the normalization 
factor $A$ is such that $g_1(r) \equiv g_c(r)$ goes to one at large distances. 
Such pair-distribution functions provide sum rules related to density- and other response functions versus density
and momentum. The solid line in the figure shows the pair-distribution function of noninteracting (NI) fermions with parallel spins:
\begin{equation}
g_{c}^{NI}(r) = 1 - \frac{9}{(k_F r)^6} \left [ \sin (k_F r) - k_F r \cos (k_F r) \right ]^2~.
\end{equation}
The $x$-axis is the interparticle distance divided with a quantity, $r_0$, which describes the average interparticle spacing:
\begin{equation}
\frac{4}{3} \pi r_0^3 = \frac{1}{\rho}~.
\end{equation}
The free Fermi gas result is close to but distinct from both QMC results (for spin-up and spin-down particles) due to the effect of
the interactions. As is to be expected, the majority component values are slightly larger than those of the minority species,
implying that it is slightly more likely to find a spin-up particle than a spin-down one.

\section{Conclusions and Future Work}

In summary, we have studied superfluid spin-polarized low-density neutron matter at small polarizations using a 
variationally optimized approach that includes the dominant well-known terms in the Hamiltonian. 
We have calculated the equation of state with the 
AV4$'$ interaction at different densities. 
We have also calculated the momentum and  pair-distribution functions. We find clear signals of a Fermi surface mismatch, as expected, and also
a linear dependence of the energy on the polarization. These results are in principle relevant to the physics
of magnetars. Furthermore, they could be tested directly by using ultracold fermionic atom gases with unequal
spin populations. In the case of cold atoms, Quantum Monte Carlo simulations of spin-polarized matter
have been used as input to density-functional theory approaches.\cite{Bulgac:2008,Bulgac:2011} Thus, our corresponding results
for neutron matter might also be used as input to self-consistent mean-field models of nuclei.

This line of Quantum Monte Carlo calculations, having first been applied to and verified in cold atomic experiments, can
also provide directions for future work in the field of nucleonic infinite matter. 
The simplest case is that of a two-component gas where the
two populations are equal.\cite{Gezerlis:2008,Gezerlis:2010} The next step is to examine the ramifications of taking
different populations for the two components: this is the case of spin-polarized low-density neutrons studied in this work.
Cold-atom experiments have by now also addressed Efimov physics, in which three components are involved. In the nuclear context,
adding a third species could provide further insight into the physics of neutron stars. If the third component particles
were taken to be protons and, as in this paper, only a few of them were added, then it would be possible to study 
highly asymmetric nuclear matter. Another possible avenue of future research is related to optical lattice experiments with cold atoms:
to first approximation these are equivalent to periodic external potentials. In the nuclear case, an external potential
would allow us to study the static response of neutron matter and would also facilitate the understanding of the impact
on neutron pairing of the ion lattice that exists in a neutron star crust.

\

\begin{acknowledgments}
The author would like to thank Joe Carlson for critically reading the manuscript
as well as Arnau Rios and Sanjay Reddy for useful discussions. 
This work was supported by DOE Grant No. DE-FG02-97ER41014.
Computations were performed at the National Energy
Research Scientific Computing Center (NERSC). 
\end{acknowledgments}


\end{document}